\documentclass{article}
\usepackage{spconf,amsmath,graphicx}
\usepackage{amssymb}
\usepackage[inline]{enumitem}
\usepackage{hyperref}
\usepackage{microtype}
\usepackage{todonotes}
\usepackage{multirow}
\usepackage{bigfoot}

\graphicspath{{./figures/}}

\newcommand\smallurl[1]{{\scriptsize \url{{#1}}}}

\title{ON USING BACKPROPAGATION FOR SPEECH TEXTURE GENERATION AND VOICE CONVERSION}
\name{Jan Chorowski, Ron J. Weiss, Rif A. Saurous, Samy Bengio}
\address{Google Brain\\{\small \texttt{\{chorowski,ronw,rif,bengio\}@google.com}}}
\begin{document}
\ninept
\maketitle
\begin{abstract}
  Inspired by recent work on neural network image generation which rely  on
  backpropagation towards the network inputs,
  we present a proof-of-concept system for speech texture synthesis and voice conversion
  based on two mechanisms: approximate inversion of the representation
  learned by a speech recognition neural network, and on matching statistics of
  neuron activations between different source and target utterances.  Similar to image
  texture synthesis and neural style transfer, the system works
  by optimizing a cost function with respect to the input waveform samples. To this end we use
  a differentiable mel-filterbank feature extraction pipeline
  and %
  train a convolutional CTC speech recognition network.
  Our system is able to extract
  speaker characteristics from very limited amounts of target speaker data,
  as little as a few seconds, and can be used to generate realistic
  speech babble or reconstruct an utterance in a different voice.
\end{abstract}
\section{Introduction}
\label{sec:intro}

Deep neural networks are a family of flexible and powerful machine
learning models. Trained discriminatively, they have become
the technique of choice in many applications, including image
recognition \cite{krizhevsky2012imagenet}, speech recognition
\cite{graves2014towards,chorowski2015attention}, and
machine translation \cite{bahdanau2014neural,sutskever2014sequence,wu2016google}.
Additionally, %
neural networks can be used to generate new data, having been applied to speech synthesis
\cite{zen2016fast,oord2016wavenet}, image generation \cite{radford2015unsupervised}, and image inpainting and
superresolution \cite{dong2014learning}.

The representation learned by a discriminatively trained deep neural network can be approximately
inverted, turning a classification model into a generator.
While exact inversion is impossible, the backpropagation algorithm can be used to find inputs which
activate the network in the desired manner. This technique has been
applied to the computer vision domain in order to gain insights into
network operation~\cite{simonyan2013deep}, %
find adversarial
examples which make imperceptible modifications to image inputs in order to change the network's
predictions~\cite{goodfellow2014explaining}, synthesize
textures~\cite{gatys2015texture}, and 
regenerate an image according to the style (essentially matching the low-level texture) of another,
referred to as \emph{style transfer}~\cite{gatys2015neural}.

In this work we investigate the possibility of converting a discriminatively
trained CTC speech recognition network into a generator. In particular, we
investigate:
\begin{enumerate*}[label=(\roman*)]
\item generating waveforms based solely on the activations of selected network
  layers, giving insights into the nature of the network's internal
  representations,
\item speech texture synthesis by generating waveforms which result in neuron
  activations in shallow layers whose statistics are similar to those of real
  speech, and
\item voice conversion, the speech analog of image style transfer, where
  the previous two methods are combined to generate waveforms which match the high
  level network activations from a \emph{content} utterance while simultaneously
  matching low level statistics computed from lower level activations from a
  \emph{style} (\emph{identity}) utterance.
\end{enumerate*}

\section{Background}

\subsection{Texture synthesis based on matching statistics}

Julesz~\cite{julesz1962visual} proposed that visual texture
discrimination %
is a function of an image's low level statistical properties.
McDermott et
al. \cite{mcdermott2009sound,mcdermott2011sound} applied the same idea to
sound, showing that perception of sound textures
relies on matching certain low level signal statistics. Furthermore,
following earlier work on image texture synthesis \cite{portilla2000parametric},
they demonstrated that simple sound textures, such as rain or fire, can be
synthesized using a gradient-based optimization procedure to iteratively update
a white noise signal to match the statistics of observed texture signals.

Recently, Gatys el al. \cite{gatys2015texture} proposed a similar statistic matching
algorithm to synthesize visual textures.  However, instead of manually
designing the relevant statistics as a function of the image pixels, they
utilized a deep convolutional neural network discriminatively trained on an image
classification task.  Specifically, they proposed to match
uncentered correlations between neuron activations in a selected
network layer.  Formally, let $C^{(n)} \in \mathbb{R}^{W
  \times H \times D}$ denote the activations of the $n$-th
convolutional layer, where $W$ is the width of the layer, $H$ is its
height, and $D$ is the number of filters. The Gram matrix of uncentered
correlations $G^{(n)} \in \mathbb{R}^{D\times D}$ is defined as:
\begin{equation}
  G^{(n)}_{i,j} = \frac{1}{WH}\sum_{w=1}^W \sum_{h=1}^H C^{(n)}_{whi}\,C^{(n)}_{whj}.
  \label{eq:image_gram_matrix}
\end{equation}

Gatys et al. demonstrated that realistic visual textures can be synthesized by matching the
Gram matrices.  In other words, the statistics necessary for texture
synthesis are the correlations between the values of two convolutional
filters taken over all the pixels in a given convolutional filter
map. We note that the Gram features in equation \eqref{eq:image_gram_matrix} are
averaged over all image pixels, and therefore are stationary
with respect to the pixel location.

\subsection{Style transfer} %

Approximate network inversions and statistic-matching texture synthesis both
generate images by minimizing a loss function with backpropagation towards
the inputs.
These two approaches can be combined to sample
images whose content is similar to a seed image, and whose texture is
similar to another one \cite{gatys2015neural}. This approach to style
transfer is attractive because it leverages a pretrained neural %
network which has learned the distribution of natural images, and therefore
does not require a large dataset at generation time -- a single image
of a given style is all that is required, and it need not be related to the
images used to train the network.

\section{Speech recognition input reconstruction}

\subsection{Network architecture}
To apply the texture generation and stylization techniques to speech
we train a fully convolutional speech recognition network following \cite{zhang2017towards}
on the Wall Street Journal dataset.
The network is trained to predict character sequences in an end-to-end fashion
using the CTC~\cite{graves2006connectionist} criterion.
We use parameters typical for a speech recognition network: 
waveforms sampled at 16kHz are segmented into 25ms windows taken every 10ms. From
each window we extract 80 log-mel filterbank features augmented with
deltas and delta-deltas. %
The 13 layer network architecture is derived from~\cite{zhang2017towards};
\begin{itemize}[noitemsep,topsep=0pt,itemindent=1em]
	\item[\bf{C0}] 128-dimensional 5$\times$5 convolution with
  2$\times$2 max-pooling,
\item[\bf{C1}] 128-dimensional 5$\times$5 convolution with 1$\times$2 max-pooling,
\item[\bf{C2}] 128-dimensional 5$\times$3 convolution,
\item[\bf{C3}] 256-dimensional 5$\times$3 convolution with 1$\times$2 max-pooling,
\item[\bf{C4-9}] six blocks of 256-dimensional 5$\times$3 convolution,
\item[\bf{FC0-1}] two 1024-dimensional fully connected layers,
\item[\bf{CTC}] a fully connected layer and CTC cost over characters,
\end{itemize}
where filter and pooling window sizes are specified in time $\times$ frequency.
All layers use batch normalization, ReLU activations, and dropout regularization.
Convolutional layers C0-9 use dropout keep probability 0.75, and
fully connected layers %
use keep probability 0.9.

The network is trained using 10 asynchronous workers with the
Adam~\cite{kingma2014adam} optimizer using $\beta_1=0.9$,
$\beta_2=0.999$, $\epsilon=10^{-6}$ and learning rate annealing from
$10^{-3}$ to $10^{-6}$. We also use L2 weight decay of $10^{-6}$.
When decoded using the extended trigram language model from the
Kaldi WSJ S5 recipe~\cite{povey2011kaldi}, the model reaches an eval
WER of $7.8\%$ on \emph{eval92}. While our network does not reach state-of-the-art
accuracy on this dataset, it has reasonable performance and is easily amenable to
backpropagation towards inputs. Even though the network was trained on the
WSJ\footnote{\smallurl{https://catalog.ldc.upenn.edu/ldc93s6a}, \smallurl{https://catalog.ldc.upenn.edu/ldc94s13a}} corpus, we
use the VCTK dataset\footnote{\smallurl{http://homepages.inf.ed.ac.uk/jyamagis/page3/page58/page58.html}}
for all subsequent experiments.

\subsection{Waveform sample reconstruction}

Our goal is to generate waveforms that will result in a particular neuron
activation pattern when processed by a deep network. Ideally,
waveform samples would be optimized directly using the backpropagation
algorithm. One possibility is to train networks that operate on raw
waveforms as in \cite{sainath2015learning}. However, it is also possible to implement the
typical speech feature pipeline in a differentiable way. We follow the
second approach, which is facilitated by readily available Tensorflow
implementation of signal processing routines \cite{abadi2016tensorflow}:
\begin{enumerate}[noitemsep,topsep=0pt]
\item Waveform framing and Hamming window application.
\item DFT computation, which multiplies waveform frames by a complex-valued DFT matrix.
\item Smooth approximate modulus computation, %
  implemented as $\text{abs}(x) \approx \sqrt{\epsilon + \text{re}(x)^2 +
    \text{im}(x)^2}$, with $\epsilon=10^{-3}$.
\item Filterbank\footnote{We use a mixed linear and mel scale, where frequencies
  below 1 kHz are copied from the STFT and higher frequencies are
  compressed using the mel scale.  We employ this scaling below
  1 kHz because the mel scale allocates too many bands to low frequencies,
  some of which are always zero when using 80 mel bands and
  256 FFT bins, which was found to be optimal for recognition.}
  feature computation, which can be implemented as a matrix multiplication.
\item Taking the elementwise logarithm of the filterbank features.
\item Computing deltas and delta-deltas using convolution over time.
\end{enumerate}

This feature extraction pipeline facilitates two methods for
reconstructing waveform samples:
\begin{enumerate*}[label=(\roman*)]
\item gradient-based optimization with backpropagation directly to the
  waveform, or
\item gradient-based optimization of the linear spectrogram,
  followed by Griffin-Lim~\cite{griffin1984signal} phase reconstruction.
\end{enumerate*}
We find that a dual strategy works best, where we
first perform spectrogram reconstruction, then invert the
spectrogram to yield an initial waveform which is further optimized directly.
We use the L-BFGS optimizer~\cite{zhu1997algorithm} for both optimization stages.

\subsection{Speech reconstruction from network activations}
\label{sec:recon}

\begin{figure}[t]
  \centering
  \setlength{\tabcolsep}{0em}
  \begin{tabular}{ccc}
    & p225 & p226 \\
    \rotatebox[origin=c]{90}{C0} &
    \raisebox{-0.5\height}[0.4\height]{\includegraphics[width=0.5\columnwidth]{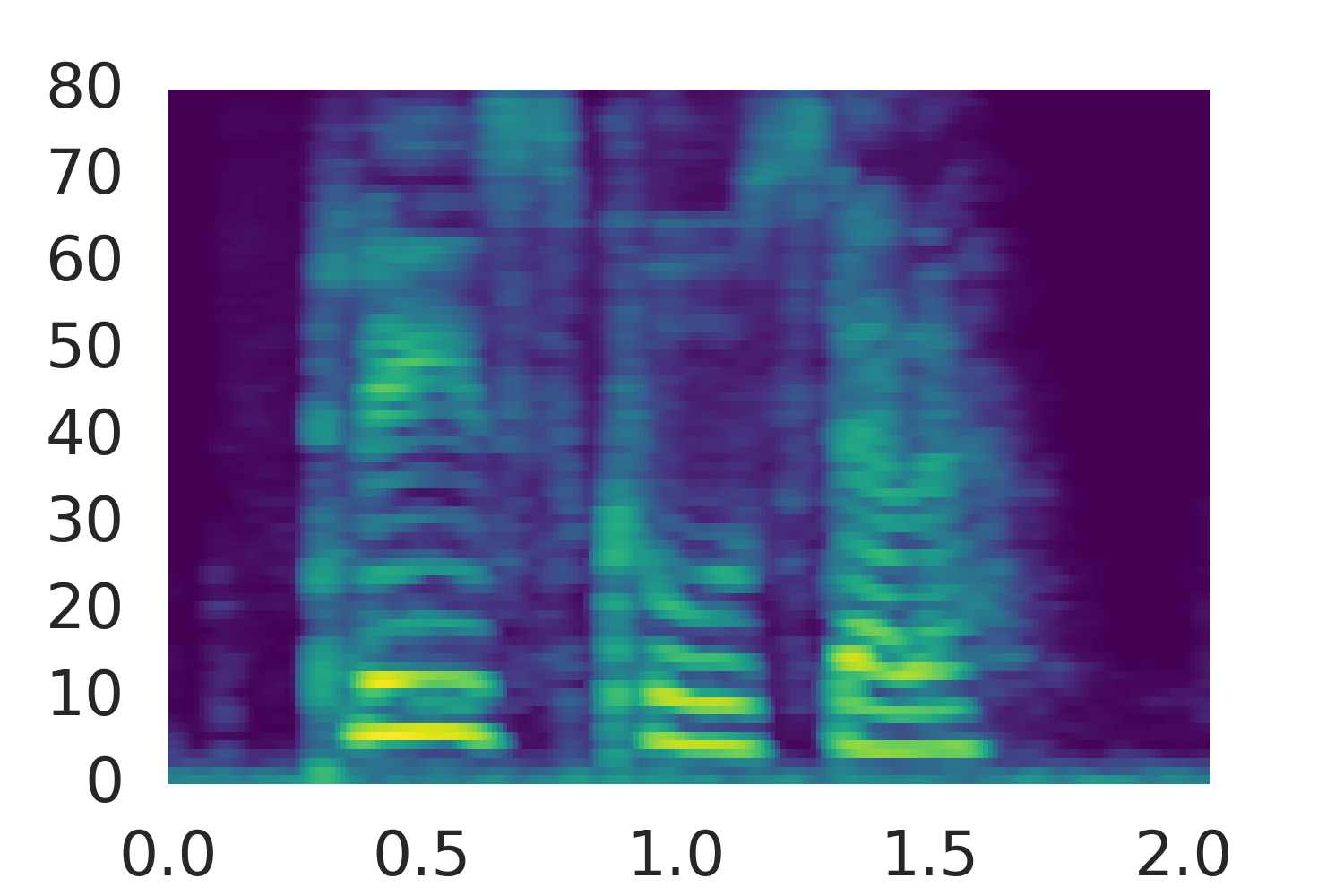}} &
    \raisebox{-0.5\height}[0.4\height]{\includegraphics[width=0.5\columnwidth]{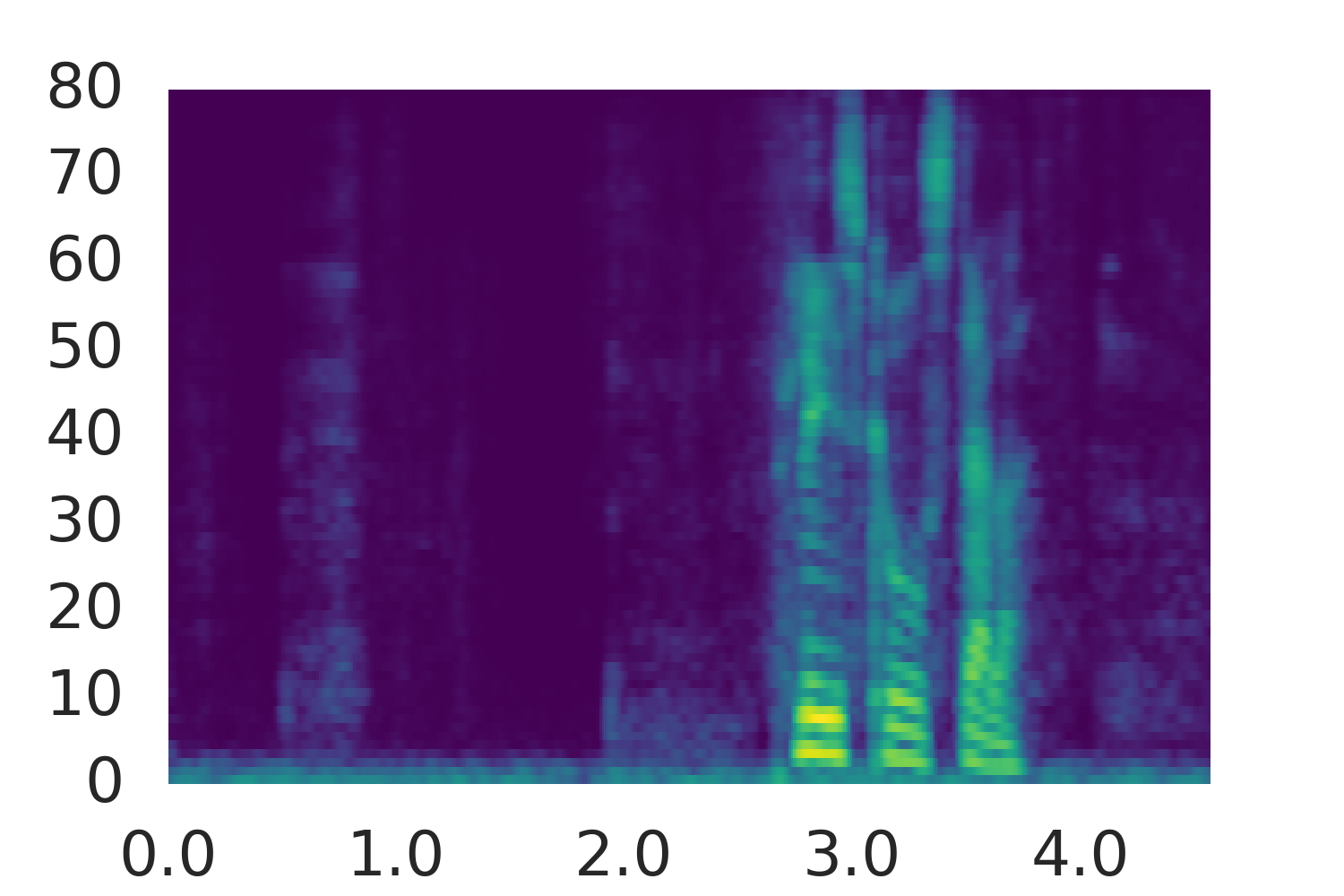}} \\
    \rotatebox[origin=c]{90}{FC1} &
    \raisebox{-0.5\height}{\includegraphics[width=0.5\columnwidth]{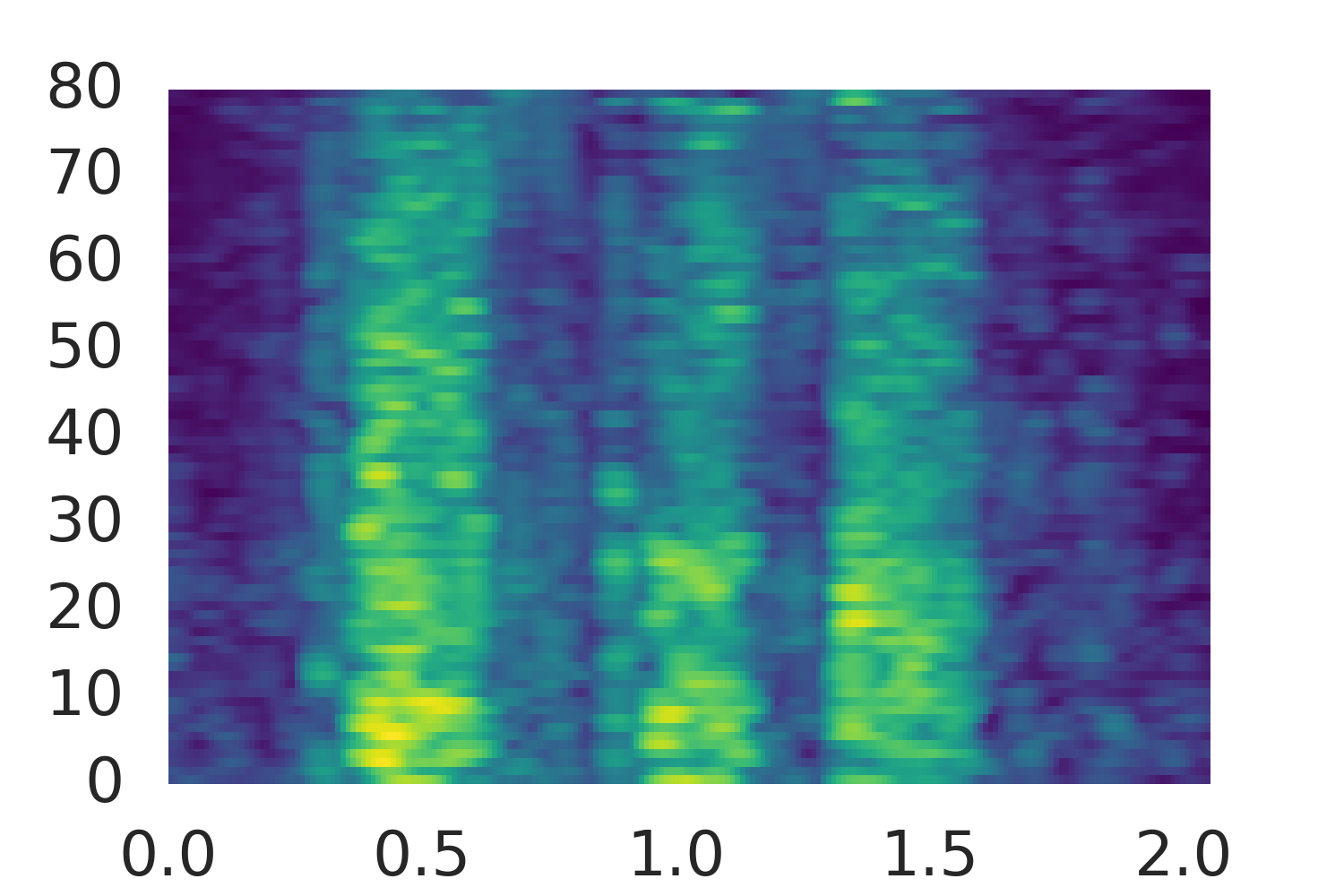}} &
    \raisebox{-0.5\height}{\includegraphics[width=0.5\columnwidth]{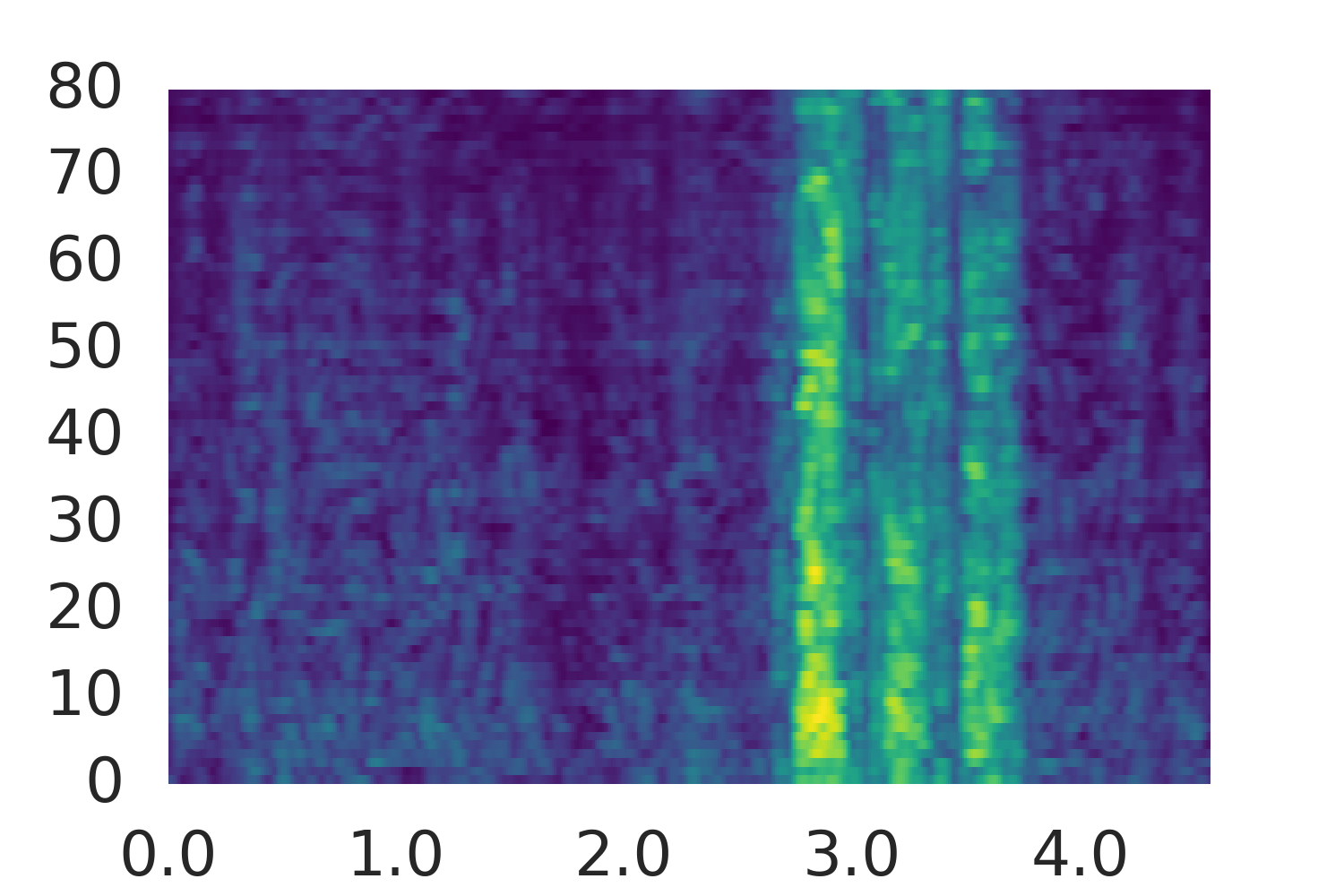}} \\
  \end{tabular}
  \caption{Mel spectrograms of waveforms reconstructed from layers C0 and
    FC1 of speakers p225 (female) and p226 (male) from the VCTK
    dataset. The reconstructions from C0 are nearly
    exact, while the reconstructions from FC1 are very noisy and are barely intelligible.
  }\label{fig:waveform_recs}
\end{figure}

We implement waveform reconstruction based on network activations following
the ReLU non-linearity in a specified layer. Figure~\ref{fig:waveform_recs} shows the spectrograms
of waveform reconstructions for speakers p225 and p226 from the VCTK
dataset
\footnote{Sound samples are available at \url{https://google.github.io/speech_style_transfer/samples.html}}.
We
have qualitatively established that waveforms reconstructed from
shallow network layers are intelligible and the speaker can be clearly
identified. Audible phase artifacts are introduced in reconstructions
from layer C3 and above, after the final pooling operation over time.
While the speech quality degrades, many speaker characteristics are
preserved in the reconstructions up to the fully connected
layers.
Listening to reconstructions from layer C9 it remains possible to
recognize the speaker's gender.

In order to reconstruct the waveforms from activations
in the fully connected layers FC0 and FC1,
we find that the reconstruction cost must be
extended with a term penalizing differences between the total energy in each feature frame of the reference and reconstruction.
We hypothesize that the network's representation in deeper layers has learned a degree of invariance to the signal magnitude, which hampers reconstruction of realistic signals.  For example, the network reliably predicts
the CTC blank symbol both for silence and white noise at different amplitudes.
The addition of this energy matching penalty enables the network to correctly reconstruct silent segments.
However, even with this additional penalty, reconstructions from layers FC0 and FC1 are highly distorted. The words are
only intelligible with difficulty and the speaker identity is lost.

\begin{figure}[t]
  \centering
  \includegraphics[width=1\columnwidth]{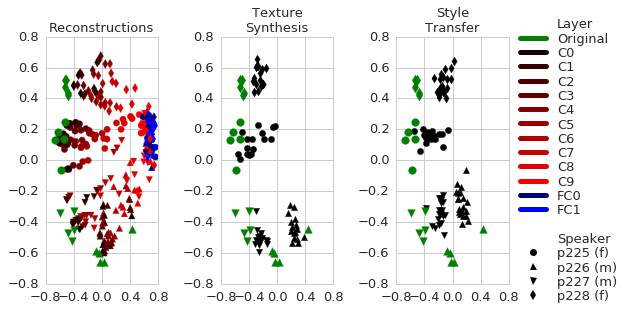}
  \caption{MDS embeddings of speaker vectors computed on original VCTK recordings,
  reconstructions from the network, synthesized waveforms, and voice converted
  waveforms. The synthesized and voice converted utterances are close to the original
  utterances and reconstructions from early layers. Reconstructions from deep layers
  converge to a single point, indicating that the speaker identity is lost.
  }\label{fig:recs_vector_embeds}
\end{figure}

To evaluate how well reconstructions based on different layers capture
characteristics of different speakers, we visualize embedding vectors
computed using an internal speaker identification system
that uses a Resnet-50 architecture \cite{hershey2017cnn}
trained on LibriVox\footnote{\smallurl{https://librivox.org/}} using a triplet-loss \cite{bredin2017tristounet}. %
Nearest neighbor classification using these embeddings obtains nearly perfect %
accuracy on the original VCTK signals.
Figure~\ref{fig:recs_vector_embeds} shows a two-dimensional MDS
\cite{kruskal1978multidimensional} embedding of these vectors.
In reconstructions from early layers, signals from each speaker cluster
together with no overlap.  As the depth increases, the embeddings for all
speakers begin to converge on a single point, indicating that the speakers
become progressively more difficult to recognize.  From this we can conclude
that the network's internal representation becomes progressively more speaker
invariant with increasing depth, a desirable property for speaker-independent speech recognition.

\subsection{Speech texture synthesis}

Unlike image textures whose statistics can be assumed to be stationary
across both spatial dimensions, the two dimensions of
speech spectrogram features, i.e. time and frequency, have different
semantics and should be treated differently.
Sound textures are stationary over time but are nonstationary across frequency.
This suggests that features extracted from
layer activations should involve correlations over time alone.  Let
$C^{(n)} \in \mathbb{R}^{T \times F \times D}$ be the tensor of activations
of the $n$-th layer of the network which consists of $D$ filters computed for $T$
frames and $F$ frequencies. The temporally stationary Gram tensor, %
$G^{(n)}\in \mathbb{R}^{F\times F\times D\times D}$,
can be written as:
\begin{equation}\label{eq:speechgram}
G^{(n)}_{ijkl} = \frac{1}{T}\sum_{t=1}^T C^{(n)}_{tik} \, C^{(n)}_{tjl}
\end{equation}

\begin{figure}[t]
  \centering
  \includegraphics[width=.8\columnwidth]{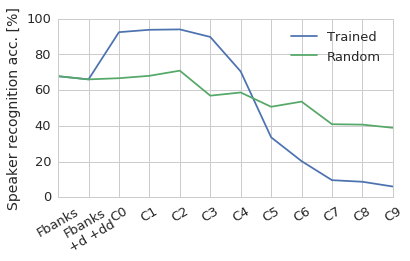}
  \caption{Accuracy of nearest-neighbor speaker classification using Gram
    tensors extracted from different network layers.}\label{fig:gram_speaker}
\end{figure}

We demonstrate that these Gram tensors capture speaker identity by using them as
features in a simple nearest neighbor speaker identification system.
Figure~\ref{fig:gram_speaker} shows speaker identification accuracy of this
system over the first 15 utterances of 30 first speakers of the VCTK dataset.
Using the lower network layers (up to C3) yields an accuracy close to
95\%, whereas using similar Gram tensors of raw mel-spectrograms extended with deltas and delta-deltas
yields only 65\% accuracy.  Deeper layers of the network
become progressively less speaker sensitive, mirroring our observations from
Figure~\ref{fig:recs_vector_embeds}.

We also observe that network training is crucial for Gram features to
become speaker-selective and for the texture synthesis to work.
After a random initialization %
the network behaves differently than it does after training: the Gram tensors
computed on shallow layers of the untrained network are less sensitive to
speaker identity than the corresponding layers in the trained network, while their
deeper layers don't exhibit as dramatic decrease in speaker sensitivity.
In contrast, image texture synthesis and style transfer have been reported to work
with randomly initialized networks \cite{he2016powerful}.

\begin{figure}[t]
  \centering
  \setlength{\tabcolsep}{0em}
  \begin{tabular}{cc}
    p225, C0 & p226, C0-C5 \\
    \raisebox{-0.5\height}[0.4\height]{\includegraphics[width=0.5\columnwidth]{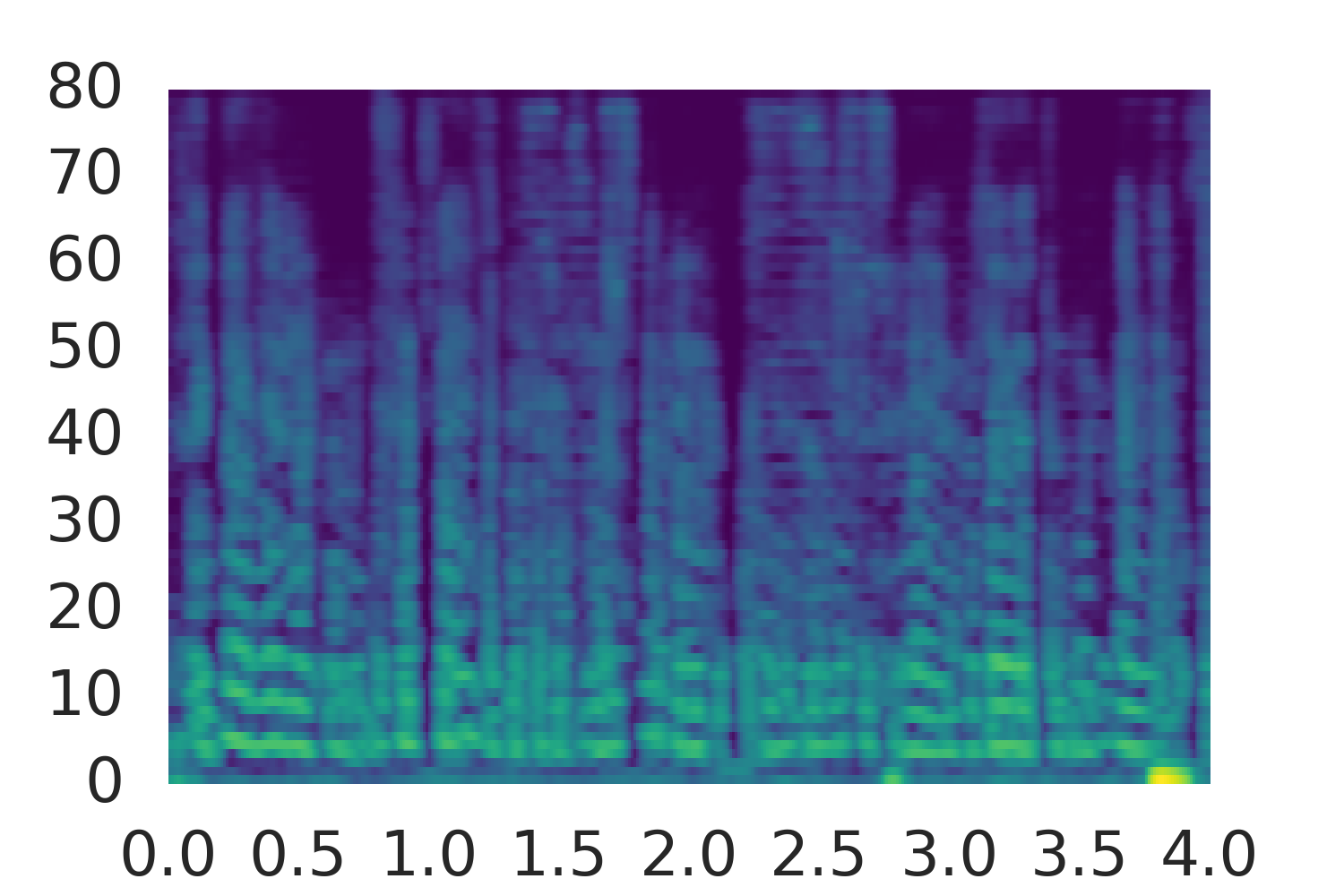}} &
    \raisebox{-0.5\height}[0.4\height]{\includegraphics[width=0.5\columnwidth]{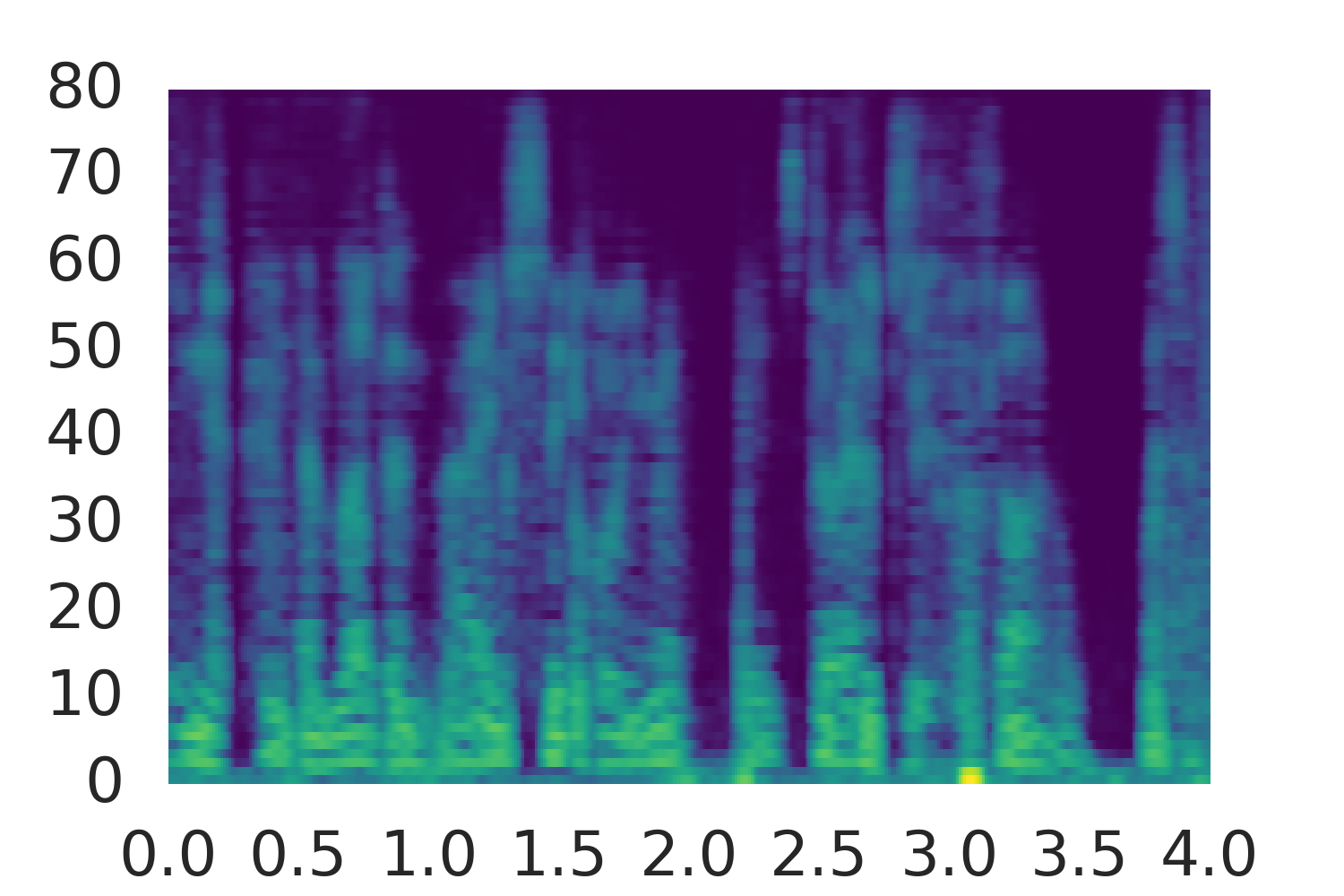}} \\
    p225, C0-C5 & Polish speaker, C0-C5 \\
    \raisebox{-0.5\height}[0.4\height]{\includegraphics[width=0.5\columnwidth]{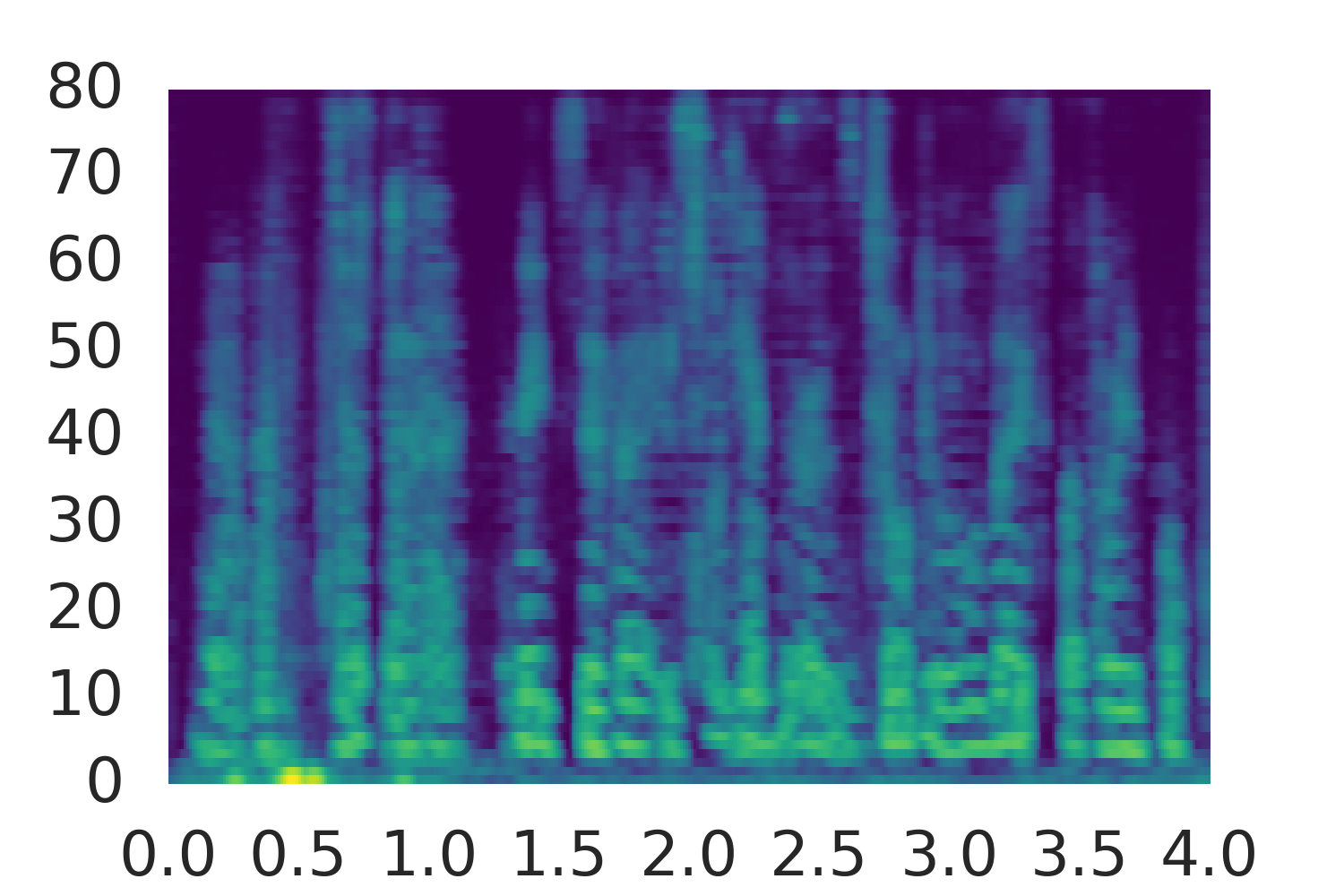}} &
    \raisebox{-0.5\height}[0.4\height]{\includegraphics[width=0.5\columnwidth]{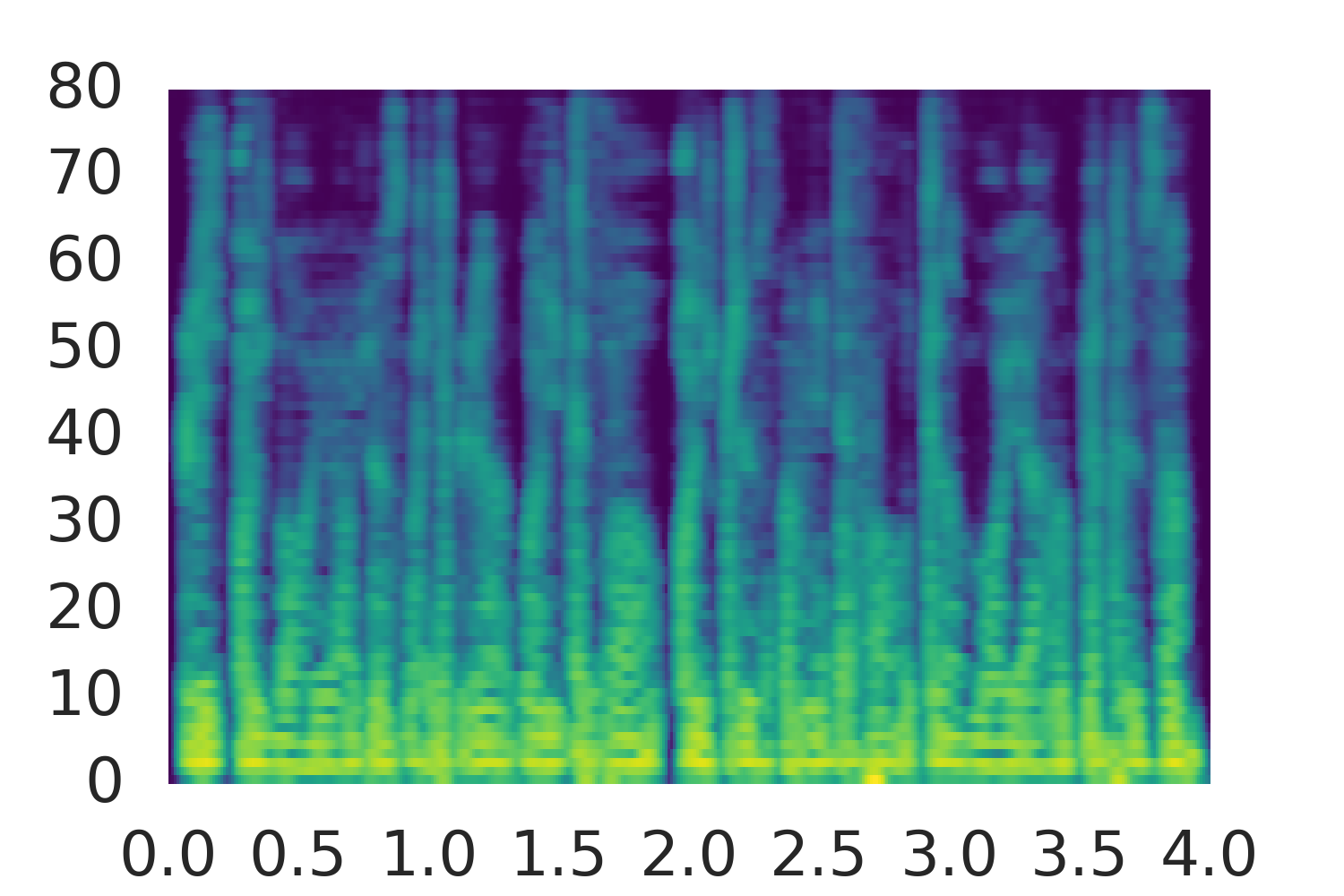}} \\
  \end{tabular}
  \caption{Mel spectrograms of textures synthesized from Gram matrices computed on 20
    utterances from VCTK speakers p225 (female) and p226
    (male), as well as a short (1s) utterance in Polish (male).
    When deeper layers
    are used, the generated sound captures more temporal
    structure. Intuitively, listening to ``p225, C0'' it is hard to
    discern words, whereas one can hear word boundaries in ``p225,
    C0-C5''. One can also see the characteristic lower pitch %
    in synthesized male voices.}\label{fig:texture_synth}
\end{figure}

Figure \ref{fig:texture_synth} shows spectrograms of generated speech
textures based on speech from the VCTK dataset and a male native Polish speaker.
The Gram tensor computed
on first layer activations captures the fundamental frequency and 
harmonics but yields a fairly uniform temporal structure. When features computed
on deeper layers are used, longer term phonemic structure can be seen,
although the overall speech is not intelligible.
This is a consequence of the increased temporal receptive field of
filters in deeper layers, where a single activation is a function of
structure spanning tens of frames, enabling the reconstruction of
realistic speech babble sounds.

\subsection{Voice conversion} %

\begin{figure}[t]
  \centering
  \setlength{\tabcolsep}{0em} 
  \begin{tabular}{ccc}
    & p225 & p226 \\ 
    \rotatebox[origin=c]{90}{Original} &
    \raisebox{-0.5\height}[0.4\height]{\includegraphics[width=0.5\columnwidth]{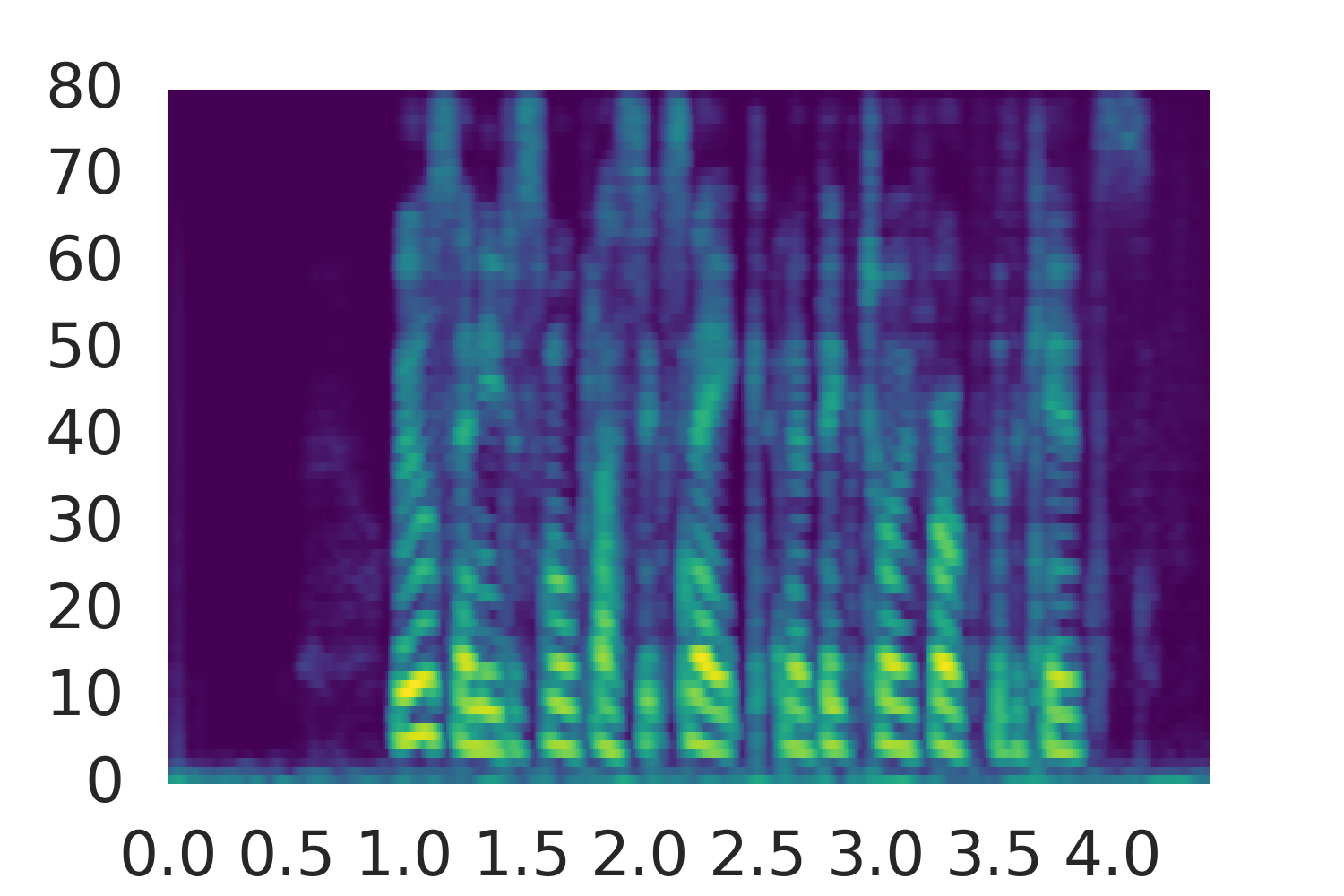}} &
    \raisebox{-0.5\height}[0.4\height]{\includegraphics[width=0.5\columnwidth]{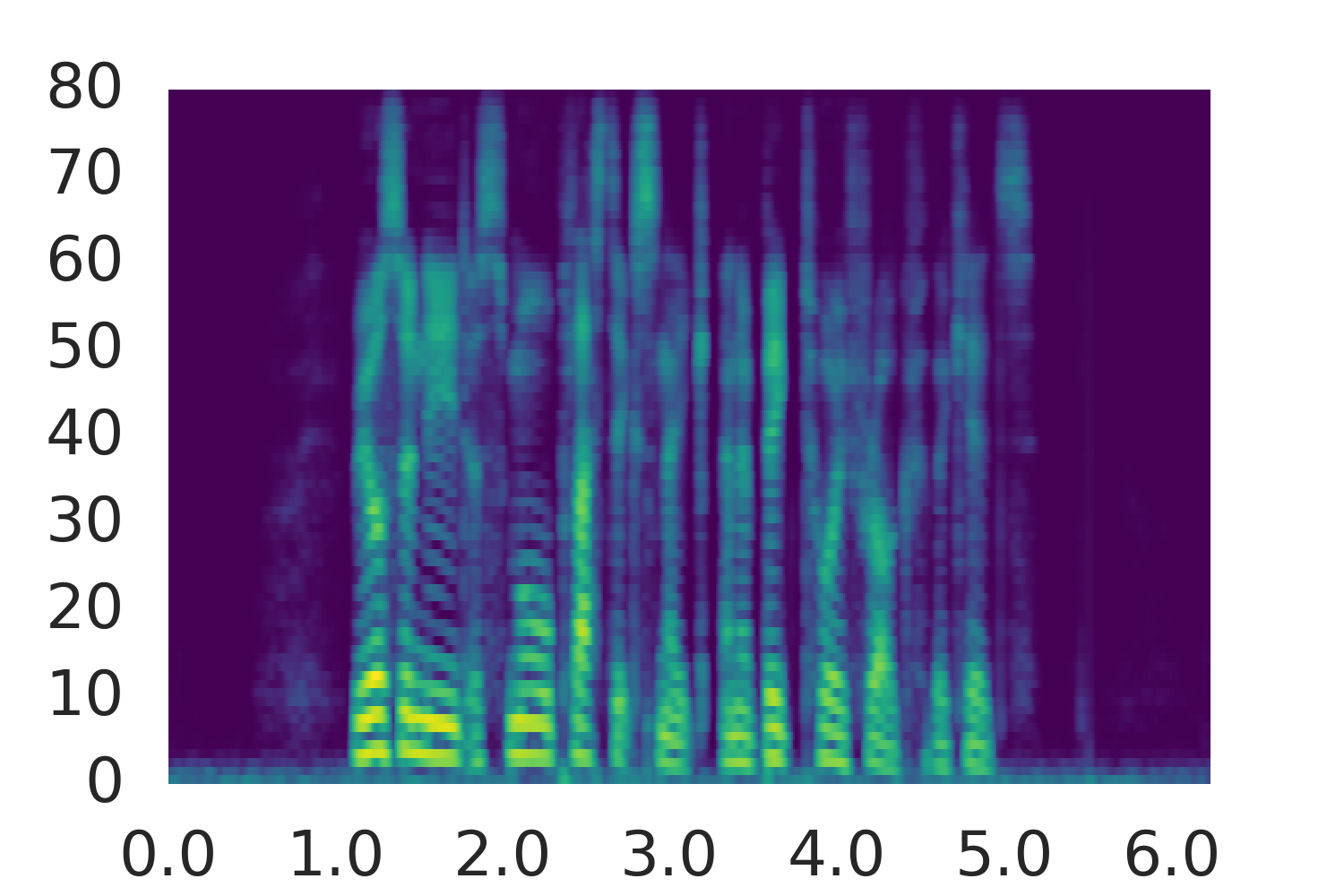}} \\
    & {p225 $\rightarrow$ p226} & {p225 $\rightarrow$ p226} \\
    \rotatebox[origin=c]{90}{Converted} &
    \raisebox{-0.5\height}[0.4\height]{\includegraphics[width=0.5\columnwidth]{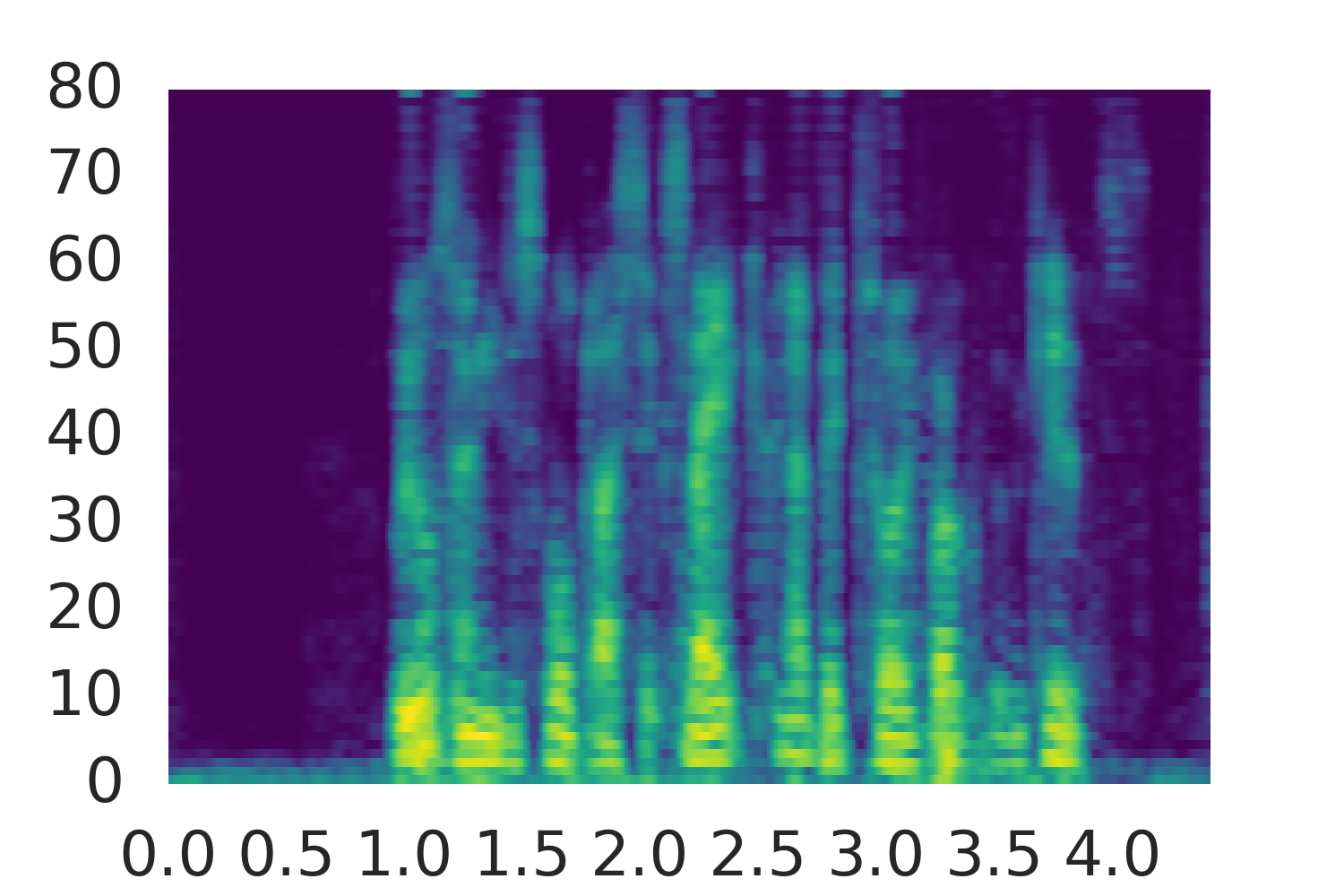}} &
    \raisebox{-0.5\height}[0.4\height]{\includegraphics[width=0.5\columnwidth]{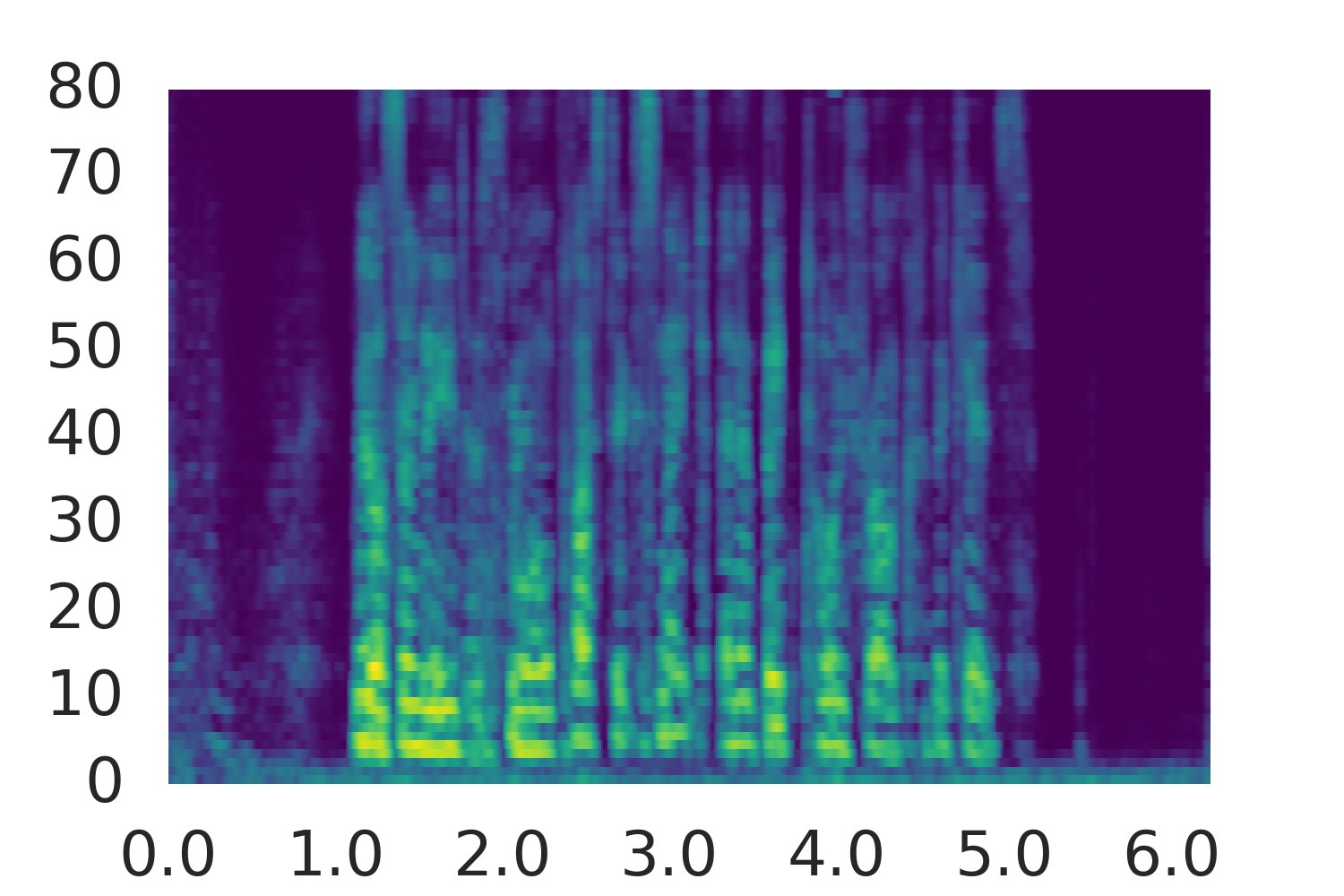}} \\
  \end{tabular}
  \caption{Mel spectrograms of voice conversion, mapping VCTK
    utterance 004 between speakers p225 (female) and p226 (male)
    performed by matching neuron activations to those from a content utterance
    and Gram features to those computed from 19 speaker identity
    utterances (about 2 minutes).}\label{fig:voice_conversion}
\end{figure}

The methods described in the previous two sections can be combined to
produce the speech analog of image style transfer: a voice conversion system.
Specifically we reconstruct the deep-layer activations of a \emph{content} utterance,
and the shallow-layer Gram features of \emph{identity} or style utterances.

Listening to the converted samples, we found that a good
tradeoff between matching the target speaker's voice and sound
quality occurs when optimizing a loss that spans all layers, with the
layers C0-C5 matched to style utterances using Gram features, and
layers C6-FC1 matched to the content utterance. We
normalize the contribution of each layer to the cost by dividing the squared
difference between the Gram or activation matrices by their dimensionality.
Furthermore, the Gram features of style layers C0-C5 use a weight of
$10^5$, activations of content layers C6-C9 use weight $0.2$ and
activations of layers FC0-FC1 use weight $10$, to base the
reconstruction on the deepest layers, but provide some signal from those in
the middle which are responsible for final voice quality.
While the speaker remains
identifiable in reconstructions from layers C6-C9 as described in Section~\ref{sec:recon}, we find that
including these layers in the content loss leads to more natural sounding synthesis.
The speaker identity still changes when the Gram feature weight
is sufficiently large.

Spectrograms of utterance generated using this procedure are shown in
Figure~\ref{fig:voice_conversion}. From the spectrograms one can see that the
converted utterances contain very different pitch, consistent with the opposite gender.
However, because the content loss is applied directly to neuron activations,
the exact temporal structure of the content utterance is retained.
This highlights a limitation of this approach: the fixed temporal
alignment to the content utterance means that it is unable to model
temporal variation characteristic to different speakers, such as changes in speaking rate.

\section{Related Work}

The success of neural image style transfer has prompted a few attempts to apply
it to audio. Roberts et al. \cite{roberts2016audio} trained audio clip embeddings using a convolutional
network applied directly to raw waveforms and attempted to generate waveforms by maximizing activations
of neurons in selected layers. The authors claim noisy results and attribute it to the low quality of the
learned filters. Ulyanov et al. \cite{ulyanov2016audio} used an untrained single-layer network to synthesize
simple audio textures such as keyboard and machine gun, and attempted audio style transfer
between different musical pieces.
The recent work of
Wyse \cite{wyse2017audio} is most similar to ours. He examines the application of
pretrained convolutional networks for image recognition and for environmental sound classification.
An example of style transfer from human speech to a crowing rooster demonstrates
the importance of using a network that has been trained on audio features, which is in line with our findings.
To the best of our knowledge, our work is the first to demonstrate
that style transfer techniques applied to speech recognition networks
can be used for voice conversion.

Speech babble sounds have been previously generated using an unconditioned WaveNet
\cite{oord2016wavenet} model trained to synthesize speech waveforms. In contrast,
we demonstrate that such complex sound textures can be generated from
a speech recognition network, using very limited amounts of data from the target speaker.

Typical voice conversion systems rely on advanced speech
representations, such as STRAIGHT \cite{kawahara1999restructuring},
and use a dedicated conversion function trained on aligned, parallel
corpora of different speakers. An
overview of the state-of-the-art in this area can be seen in the recent Voice
Conversion Challenge \cite{toda2016voice}.
While our system produces samples that have an
inferior quality, it operates using a different and novel principle:
rather than learning a frame-to-frame conversion, it uses a speech
recognition network to define a speaker similarity cost %
that can be optimized to change the perceived identity of the
speaker.

\section{Limitations and Future Work}
We demonstrate a proof-of-concept speech texture synthesis and voice conversion
system that derives a statistical description of the target voice from the activations
of a deep convolutional neural network trained to perform speech recognition.
The main benefit of the proposed approach %
is the ability to utilize very limited amounts of data from the target speaker.
Leveraging the distribution of natural speech captured by the pretrained network, 
a few seconds of speech are sufficient to synthesize recognizable
characteristics of the target voice.
However, the proposed approach is also quite slow, requiring several thousand
gradient descent steps. In addition, the synthesized utterances are of relatively low
quality.

The proposed approach can be extended in may ways. First, analogously to the
fast image style transfer algorithms 
\cite{ulyanov2016texture,johnson2016perceptual,dumoulin2016learned},
the Gram tensor loss can be used as additional
supervision for a speech synthesis neural network such as WaveNet \cite{oord2016wavenet} or
Tacotron \cite{wang2017tacotron}.
For example, it might be feasible to use the style loss to
extend a neural speech synthesis system to a wide set of speakers given only
a few seconds of recorded speech from each one.
Second, the method depends on a pretrained speech recognition
network.  In this work we used a fairly basic network using feature
extraction parameters tuned for speech recognition. Synthesis quality could probably be improved
by using higher sampling rates, increasing the window overlap and running the network on
linear-, rather than mel-filterbank features.

\section{Acknowledgments}
Authors thank Yoram Singer, Colin Raffel, Matt Hoffman, Joseph Antognini, and Navdeep Jaitly for helpful discussions and inspirations,
RJ Skerry-Ryan for signal processing in TF, and Aren Jansen and Sourish Chaudhuri
for help with the speaker identification system.

\bibliographystyle{IEEEbib}
\bibliography{refs}

\end{document}